\documentclass[aps,prx,reprint,superscriptaddress,longbibliography,floatfix]{revtex4-2}

\usepackage{graphicx}
\usepackage{amsmath,amssymb}
\usepackage{bm}
\usepackage{booktabs}
\usepackage{url}
\usepackage{xcolor}
\graphicspath{{figs/}}

\begin{document}

\title{Extrapolating the emergence of Hamiltonian chaos with random-feature Hamiltonian neural networks}

\author{Jaesung Choi}
\email[Author to whom correspondence should be addressed: ]{joseph9463@kias.re.kr}
\affiliation{Center for Artificial Intelligence and Natural Sciences, Korea Institute for Advanced Study, Seoul 02455, Korea}

\date{\today}

\begin{abstract}
Machine learning of Hamiltonian dynamics has driven growing interest in Hamiltonian neural
networks (HNNs), which encode Hamilton's equations of motion into the learning architecture.
Despite this progress, it remains unknown whether such networks can predict dynamical regimes
absent from their training data, in particular the broad chaotic sea that emerges beyond
the observed parameter interval. We address this question using a parameter-aware
random-feature Hamiltonian neural network (RF-HNN). Trained using data from only a small
number of control-parameter values at which invariant tori dominate, the RF-HNN predicts
autonomous long-time dynamics at unseen parameter values where mixed phase space develops
and chaotic regions expand, with no data from that regime used in training or model
selection. The method is demonstrated across four two-degree-of-freedom Hamiltonian
families, including the H\'enon--Heiles system. Using Poincar\'e-section geometry and
finite-time Lyapunov exponents, we show that the RF-HNN reproduces the breakup of regular
structures and the emergence and growth of chaotic regions, whereas conventionally
trained HNNs with the same Hamiltonian structure remain too regular. These results show that what decides
parameter extrapolation is not Hamiltonian structure alone but how the fitted Hamiltonian
continues in the control parameter. To our knowledge,
this is the first demonstration that a learned Hamiltonian can qualitatively extrapolate
from predominantly regular dynamics into a broad chaotic sea absent from training.
\end{abstract}

\maketitle

\begin{figure*}[t]
\centering
\includegraphics[width=\linewidth]{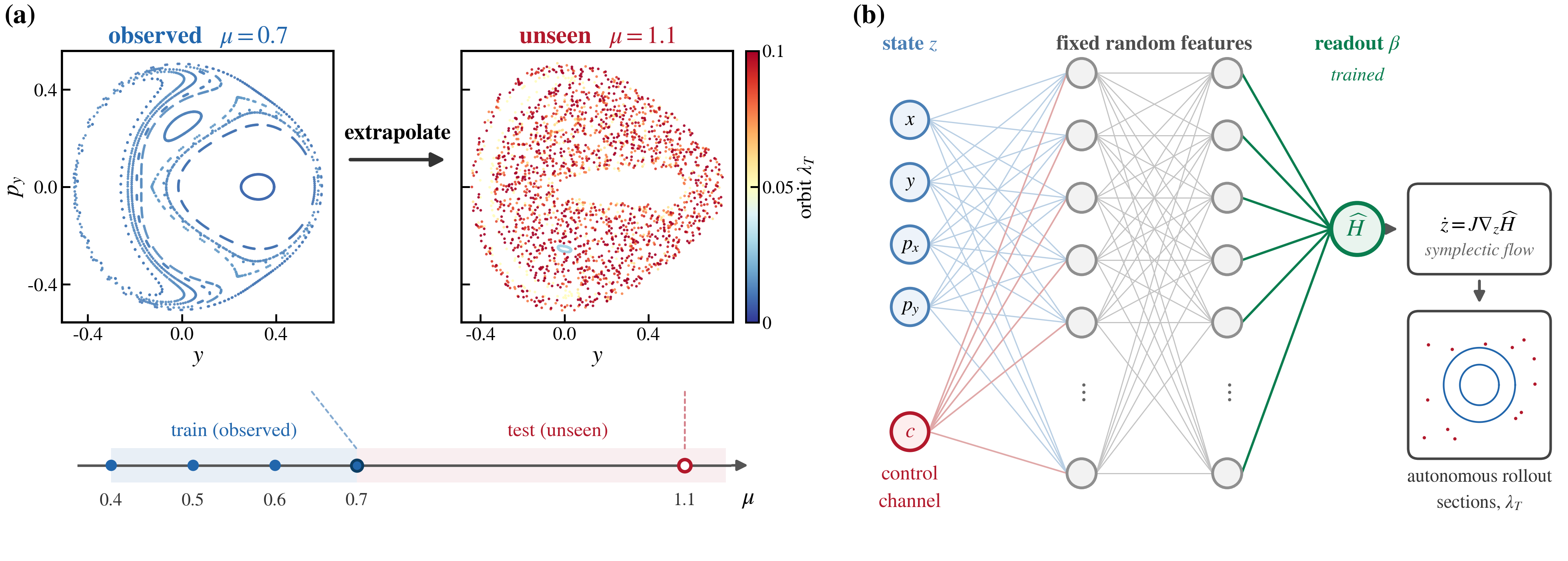}
\caption{Problem setting and model. (a) Truth Poincar\'e sections of the canonical
linear-control H\'enon--Heiles family at an observed training control and at the unseen
target control, with every orbit colored by its own finite-time Lyapunov exponent (navy =
lower and red = higher finite-time instability). Each section pools 200 crossings from each
of 12 orbits. The
parameter axis below marks the training interval with its four observed controls (filled dots)
and the unseen test control (open dot).
(b) The parameter-aware RF-HNN. The state $z$ (blue) and the control $c$ (red), the latter supplied through a dedicated parameter channel, pass through two fixed random-feature layers sampled
once and never trained. The readout $\beta$ (green) is the only fitted part of the model,
obtained by ridge regression. Hamilton's equations applied to $\widehat{H}$ define the
autonomous flow, sketched here as a Poincar\'e section with regular orbits (blue curves)
and chaotic crossings (red points).}
\label{fig:setup}
\end{figure*}

\section{Introduction}\label{sec:intro}

Near-integrable Hamiltonian systems exhibit one of the most thoroughly studied routes to
chaos in nonlinear dynamics. As a perturbation or control parameter grows,
Kolmogorov--Arnold--Moser (KAM) tori are progressively destroyed, resonance layers widen and
overlap, and an initially thin chaotic layer grows into a sea that dominates the energy
shell \cite{chirikov1979,lichtenberg1992}. The resulting mixed phase space, in which islands
of regular motion coexist with chaotic regions, is generic for conservative systems ranging
from galactic potentials to molecular vibrations \cite{henon1964,barbanis1966,kryvohuz2010}.
On a Poincar\'e section, the progression appears as the gradual breakup of invariant curves
and the spread of scattered points over the accessible area, as exemplified by the
H\'enon--Heiles system \cite{henon1964}. A one-parameter family of Hamiltonians therefore
spans the full range of behavior from predominantly regular to strongly chaotic motion.

Reconstructing such dynamics from data becomes necessary when the governing Hamiltonian is
not known in explicit form. Conventional neural networks handle this task poorly, as a model
trained directly on trajectory data does not inherit the conservation laws of the underlying
flow, and its long-time predictions acquire spurious dissipation or energy drift even when
training data are abundant \cite{greydanus2019hnn}. Hamiltonian neural networks (HNNs)
address this failure by representing a scalar function $H_\theta$ and obtaining the vector
field from Hamilton's equations, so that the learned flow conserves the learned energy and
phase-space volume by construction \cite{greydanus2019hnn}. Symplectic architectures take a
complementary route and learn the discrete-time flow map itself, either by composing
elementary symplectic modules or through parameterizations that are symplectic by
construction \cite{jin2020sympnets,chen2021gfnn}. Structure-preserving models of this kind
have delivered stable long-time predictions for systems ranging from simple oscillators to
many-body problems and have conserved energy over horizons where conventional networks drift \cite{greydanus2019hnn,jin2020sympnets,chen2021gfnn}. Trained across the
order-to-chaos transition of the H\'enon--Heiles system, HNNs also reproduce both regular
and chaotic behavior, at conditions covered by the training data \cite{choudhary2020order}. The generalization has been pushed even
further by adaptable HNNs, which take the control parameter as an additional input and have
reconstructed the Hamiltonian dynamics of a family at control values absent from the
training set \cite{han2021adaptable}. In these demonstrations, however, the training data already
contained chaotic orbits, so predictions at new control values, including values outside
the training set, reproduced dynamical regimes that were represented in training rather
than a qualitatively new one.

Extrapolation beyond the training range is a far more demanding task \cite{goring2024ood},
but a series of recent results in reservoir computing has shown that learned models can
cross dynamical regimes \cite{kim2021teaching,kong2021collapse,choi2022ews}. In
this approach, a fixed, randomly generated recurrent network is driven by the observed time
series, and its internal states serve as random features on which only a linear readout is
trained \cite{jaeger2004harnessing,choi2025homotopy}. Parameter-aware reservoir
computers that receive the control as an additional input
have reconstructed attractors at controls not sampled in training \cite{kim2021teaching},
predicted critical transitions and system collapse from pre-transition data \cite{kong2021collapse,panahi2024adaptable}, and extrapolated tipping points of
non-stationary systems
when internal hyperparameters such as the spectral radius are optimized \cite{koglmayr2024tipping}. Regime-crossing results continue to accumulate in other
settings: parameter-dependent neural ordinary differential equations extrapolate across
bifurcations of dissipative systems \cite{tegelen2025bifurcations}, and an evolution-operator
emulator trained only on chaotic dynamics predicts an unseen laminar regime \cite{shokar2025outoftraining}.
Extrapolation of Hamiltonian dynamics, however, remains largely unexplored. Reservoir
computing has been applied to Hamiltonian systems, reconstructing KAM diagrams from
trajectories recorded at a few parameter labels \cite{zhang2021rc}. In the mixed-regime
demonstrations, however, chaotic motions were part of the training data, and the authors
note that quasi-periodic training motions alone do not suffice to replicate the chaotic
dynamics. The framework also offers no mechanism that enforces Hamiltonian structure by
construction. Yet the
random-feature strategy underlying reservoir computing is not tied to recurrent networks.
Random-feature Hamiltonian models replace the reservoir with a static set of nonlinear
features, drawn at random or sampled from data, and use the readout to represent the
Hamiltonian itself, so that the fit reduces to a ridge regression solved in closed form \cite{jakovac2022,bolager2023swim,rahma2024backpropfree,rahma2026hgn}. Models of this type
retain the conservation structure of an HNN while avoiding back-propagation entirely, and
for a single training system, data-driven sampling schemes such as SWIM have reached
accuracies comparable to those of back-propagated networks at a fraction of the training cost \cite{bolager2023swim,rahma2024backpropfree,rahma2026hgn}.

\begin{table*}[t]
\caption{Final systems and evaluation ranges. Each listed training value supplies an
independently sampled energy-shell data set. Held-out interpolation sections are reported
for the bounded family. All extrapolation controls lie strictly above the corresponding training interval. $\Delta t$ is the integration time step.}
\label{tab:systems}
\centering
\small
\setlength{\tabcolsep}{5pt}
\begin{tabular}{@{}lccccc@{}}
\toprule
Family & $E_0$ & Training controls & Interpolation controls & Extrapolation controls & $\Delta t$ \\
\midrule
Linear-$\mu$ H\'enon--Heiles
& 0.13 & $\{0.4,0.5,0.6,0.7\}$ & ---
& $\{0.80,0.95,1.10\}$ & 0.05 \\
Bounded $\mu^2$ H\'enon--Heiles
& 0.16 & $\{0.9,1.0,1.1,1.2\}$ & $\{0.95,1.05,1.15\}$
& $\{1.30,1.40,1.50\}$ & 0.05 \\
Confined Barbanis-type
& 0.18 & $\{0.8,1.0,1.2\}$ & ---
& $\{1.40,1.60\}$ & 0.04 \\
Bilinear Morse pair
& 0.18 & $\{0.3,0.4,0.5\}$ & ---
& $\{0.60,0.70,0.80,1.10\}$ & 0.05 \\
\bottomrule
\end{tabular}
\end{table*}

In this study, we apply this random-feature construction to the extrapolation of a
Hamiltonian transition. Instead of drawing the features once and using them as given, we
optimize internal hyperparameters such as the feature scales, as in reservoir
computing, and the fast training provided by the single closed-form ridge fit is what
makes this optimization affordable. We show that the resulting parameter-aware
random-feature HNN (RF-HNN; Fig.~\ref{fig:setup}), trained only at control values where
regular motion dominates, extrapolates the emergence of widespread chaos beyond the training
range. The extrapolation regime contributes nothing to training or model selection,
and the test is carried out on four two-degree-of-freedom families, namely two
H\'enon--Heiles families, a confined Barbanis-type model, and a bilinearly coupled Morse
pair \cite{henon1964,barbanis1966,morse1929}, against conventionally trained parameter-aware HNNs \cite{greydanus2019hnn,han2021adaptable}, both as a single network and as a 20-network ensemble. Because every model in the comparison is Hamiltonian by
construction, differences in the extrapolated dynamics can be attributed to the fitted
continuation in the control rather than to the presence of conservation structure. In
all four families, the RF-HNN tracks the growth of the chaotic sea quantitatively, whereas
the conventionally trained models, including the ensemble, fail to track that growth and remain too
regular at the largest extrapolation controls. To our knowledge, no learned Hamiltonian has previously crossed this transition from a
regular-dominated training regime, and the comparison identifies parameter continuation,
rather than conservation structure alone, as the decisive factor.

\section{The regular-to-chaotic extrapolation task}
\label{sec:task}

We consider one-parameter families of two-degree-of-freedom Hamiltonians
$H_c(z)$ with $z=(x,y,p_x,p_y)$, in which increasing the control $c$ drives
the standard near-integrable route to chaos. We write $H(z,c) \equiv H_c(z)$ when the control appears as an explicit argument. Along this route, invariant
tori break up, resonance layers overlap, and a chaotic sea spreads over the
energy shell.
Figure~\ref{fig:setup}(a) illustrates the extrapolation problem posed in
this work for the canonical H\'enon--Heiles family of
Sec.~\ref{sec:canonicalHH}. At the observed control, the Poincar\'e section
is organized by invariant curves and the orbit-wise finite-time Lyapunov
exponents remain uniformly small. At the unseen target control, the same
sampling yields a section dominated by a chaotic sea. The
parameter axis beneath the sections summarizes the data constraints of the
task. The learner receives vector-field samples
$f_i = J\nabla_z H(z_i, c_i)$, where $J$ is the canonical symplectic matrix, on a fixed energy shell $E_0$, drawn only at the filled training controls, at which the section dynamics is dominated by
invariant tori. Model selection, including all hyperparameter tuning, is
likewise restricted to controls at or inside the training interval. The
trained model is then integrated autonomously at the open test control,
which lies strictly above the training interval and at which the true
family develops a broad mixed phase space. Success is judged not by
short-time trajectory error but by the long-time dynamical content of the
learned flow, namely the occupied area and geometry of its Poincar\'e
sections and its orbit-wise finite-time Lyapunov exponents, compared with
truth by identical procedures (Secs.~\ref{sec:poincare}
and~\ref{sec:lyapunov}). No data from the extrapolation regime are
used for training or model selection at any stage.

The two panels of Fig.~\ref{fig:setup}(a) make clear that this task differs
from interpolation in kind rather than in degree. Within the training interval,
a faithful model needs only to connect regular behavior of the kind it has
already observed. Beyond this interval, it must produce a qualitatively new form
of long-time behavior from a fitted Hamiltonian whose continuation in $c$
was never supervised. This is the transition from the left panel of
Fig.~\ref{fig:setup}(a) to the right. Every model considered below is
Hamiltonian by construction, but nothing in that constraint determines how
the fitted Hamiltonian behaves at controls beyond the training interval. A
model can conserve its learned energy exactly and still predict regular
motion where the true family has become chaotic.

We instantiate the task on four families, summarized in
Table~\ref{tab:systems}. Two H\'enon--Heiles families carry the principal
claims, and two auxiliary families test whether the results generalize. We report a dedicated
interpolation test for the bounded family. Every control designated for
extrapolation lies strictly above its family's training interval.

\subsection{Canonical H\'enon--Heiles family}
\label{sec:canonicalHH}

The canonical family is
\begin{equation}
\begin{aligned}
H^{\mathrm{HH}}_{\mu}(z)
&= \tfrac{1}{2}\bigl(p_x^2+p_y^2\bigr)
 + \tfrac{1}{2}\bigl(x^2+y^2\bigr) \\
&\quad + \mu\Bigl(x^2 y - \tfrac{y^3}{3}\Bigr),
\end{aligned}
\label{eq:HH}
\end{equation}
where $\mu=1$ recovers the standard H\'enon--Heiles
Hamiltonian \cite{henon1964}. We fix $E_0=0.13$ and vary $\mu$.
Rescaling $z \mapsto \mu z$ maps the family at energy $E_0$ onto the
standard system at energy $\mu^2 E_0$, so raising $\mu$ is equivalent to raising the energy of the standard system. Training uses
$\mu = 0.4, 0.5, 0.6, 0.7$, for which the Poincar\'e section is still
dominated by invariant curves, and extrapolation extends to $\mu = 1.1$,
at which the true section develops a broad chaotic component. The escape
channels of the potential open at $E_{\mathrm{esc}} = 1/(6\mu^2)$. At
$\mu = 1.1$ the shell energy $E_0$ still lies below this threshold, but
the margin is small and closes near $\mu = 1.13$, so the canonical
potential does not admit a substantially deeper bounded extrapolation.
Going beyond this limit requires modifying the potential itself, which
is the step taken in Sec.~\ref{sec:boundedHH}.

\subsection{Bounded nonlinear-control family}
\label{sec:boundedHH}

The more demanding test is
\begin{equation}
\begin{aligned}
H^{\mathrm{BHH}}_{\mu}(z)
&= \tfrac{1}{2}\bigl(p_x^2+p_y^2\bigr)
 + \tfrac{1}{2}\bigl(x^2+y^2\bigr) \\
&\quad + \mu^2\Bigl(x^2 y - \tfrac{y^3}{3}\Bigr)
 + \kappa\bigl(x^2+y^2\bigr)^2,
\end{aligned}
\label{eq:BHH}
\end{equation}
with $\kappa = 0.5$ and $E_0 = 0.16$. Training controls are
$\mu = 0.9, 1.0, 1.1, 1.2$ and extrapolation reaches $\mu = 1.5$. The two
modifications play distinct roles. The square on $\mu$ tests nonlinear
parameter continuation. For any fixed control it only reparameterizes the
cubic coefficient, but the network receives $\mu$ rather than $\mu^2$ and
must continue this dependence from data. The quartic term genuinely
changes the Hamiltonian and confines the motion, converting escape into
bounded chaotic transport and thereby permitting an extrapolation deeper
into the chaotic regime than Eq.~\eqref{eq:HH} allows. We accordingly use Eq.~\eqref{eq:BHH} as a
controlled, more demanding H\'enon--Heiles-type test rather than as a claim about
the standard system.

The two auxiliary families, a quartically confined Barbanis-type model and
a bilinearly coupled Morse pair, are defined in Appendix~\ref{app:systems}
and are used only in Sec.~\ref{sec:generalization}, which tests whether
the extrapolation generalizes across Hamiltonians. Across all four
families we describe the training regimes as predominantly regular rather
than strictly regular. None of the training sections contains a broad
chaotic sea, although at the upper training controls of the two
H\'enon--Heiles families a minority of sampled orbits shows mildly
elevated finite-time instability.

\section{Models and numerical protocol}
\label{sec:models}

Let $c$ denote the control parameter of a given family. Our model, the parameter-aware RF-HNN shown in Fig.~\ref{fig:setup}(b), represents the Hamiltonian of the family as
\begin{equation}
\begin{aligned}
\phi_1(z,c)&=\sigma(W_z z+w_c c+b_1),\\
\phi_2(z,c)&=\sigma(W\phi_1(z,c)+b_2),\\
\widehat{H}_{\beta}(z, c) &= \beta^{\mathsf{T}}\phi_2(z, c),
\end{aligned}
\label{eq:rf}
\end{equation}
where $\sigma$ is an elementwise activation function and the state weights $W_z$, the control weights $w_c$, the second-layer matrix $W$, and the biases $b_1$ and $b_2$ are sampled once and kept fixed. Only the
readout $\beta$ is trained. For a training state $z_i$ with target field
$f_i = J\nabla_z H(z_i, c_i)$,
\begin{equation}
J\nabla_z \widehat{H}_{\beta}(z_i, c_i) = A_i \beta,
\qquad
A_i = J\,\bigl[\partial_z \phi_2(z_i, c_i)\bigr]^{\mathsf{T}},
\label{eq:linmap}
\end{equation}
with the feature Jacobian $\partial_z\phi_2$ evaluated in closed form
through the fixed layers,
and stacking the $A_i$ and $f_i$ reduces the fit to a ridge regression,
\begin{equation}
\beta = \bigl(A^{\mathsf{T}}A + \alpha I\bigr)^{-1} A^{\mathsf{T}} f,
\label{eq:ridge}
\end{equation}
with regularization parameter $\alpha$.
For fixed features and hyperparameters, the readout fit is a convex
least-squares problem, and Eq.~\eqref{eq:ridge} is its unique solution, with no iterative optimization involved. Hamiltonian values are not
required for training, and the additive constant of $\widehat{H}_{\beta}$
left undetermined by field matching does not affect the vector field. Hamilton's equations applied to $\widehat{H}_{\beta}$ then
define the autonomous flow used for all rollouts and diagnostics.

The scales of the random features act as internal hyperparameters and,
as in reservoir computing, are optimized rather than fixed in advance. The state weights $W_z$ and control weights $w_c$ are independent
standard normal draws multiplied by $s_z$ and $s_c$, respectively. The
second-layer matrix $W$ is a standard normal draw multiplied by $g_2/\sqrt{N}$, where $N$ is the common width of the two feature layers. The factor $1/\sqrt{N}$ offsets the $N$-term sums in
$W\phi_1$, so that $g_2$ sets the second-layer scale independently of
the width. Every bias component is drawn independently and uniformly from $[-1,1]$. For each of the three candidate activations ($\tanh$, $\mathrm{GELU}$,
and $\mathrm{softplus}$), 100 Bayesian-optimization trials independently
optimize $N$, $s_z$, $s_c$, $g_2$, and the regularization parameter
$\alpha$ \cite{optuna2019}. Each trial fits the readout at a
designated subset of the training controls and is scored by the
root-mean-square error (RMSE) of the predicted vector field on an
independently sampled shell at one validation control inside the
training interval.
The activation and all remaining hyperparameters are then selected by
that validation RMSE. No extrapolation control is used in model
selection. Because each trial costs only one closed-form ridge fit, the
300 trials per family remain affordable. After selection, the readout is
refitted at all training controls using 8000 independently sampled shell
states per control, and reported results average three
random-feature seeds. Full details of the search and the
selected configurations are given in Appendix~\ref{app:hpo}.

Training states are sampled independently rather than extracted from
trajectories. We draw $(x,y)$ uniformly from a family-specific rectangular
proposal box and reject any point with potential energy $V(x,y;c) \geq E_0$, for which the shell leaves no kinetic energy. At each accepted point, the remaining
energy $E_0 - V$ fixes the momentum magnitude and leaves only its
direction free: an angle $\theta$ drawn uniformly on $[0, 2\pi)$ gives
$(p_x, p_y) = [2(E_0 - V)]^{1/2}(\cos\theta, \sin\theta)$, which places
the state on the designated energy shell. Appendix~\ref{app:systems}
gives the proposal boxes.

For comparison, we train a conventional HNN in which every weight is
trained end to end by back-propagation on all training controls,
hereafter the MLP-HNN: two
hidden $\tanh$ layers of width 200 map $(z, c)$ to a scalar
$H_{\theta}$, following the standard parameter-aware
design \cite{greydanus2019hnn,han2021adaptable}. The baseline receives
the same field-matching objective and states from the same designated
energy shell as the RF-HNN, and its
architecture and training budget are fixed without consulting any
extrapolation control. We evaluate both single networks and a 20-network ensemble, whose prediction is the gradient of the mean of
the learned Hamiltonians. This is a controlled comparison of fitting methods on data drawn from
the same shells and controls. The training details are given in
Appendix~\ref{app:systems}.

\subsection{Poincar\'e diagnostics}
\label{sec:poincare}

At every control, 28 initial conditions are sampled from the section
$x = 0$, $p_x > 0$ on the true energy shell and then shared by truth and
all learned models. Section crossings are linearly interpolated to
$x = 0$. We retain exactly 50 crossings per orbit. The one exception is the
confined Barbanis-type potential, where the broken $x \to -x$ symmetry
makes some orbits cross the section several times more often than
others. There we retain 20 crossings, which keeps the sample balanced
across orbits. Every trajectory evaluated in the main comparison reaches its target count and none escapes.

The Poincar\'e sections displayed in the figures use a denser,
visualization-only cloud: the first 12 of the same 28 initial conditions,
integrated to 200 crossings per orbit so that invariant curves appear
continuous at print size. All reported statistics use the 28-orbit
sample above.

We use two complementary section metrics. First, the occupancy, a
coarse-grained measure of the occupied section area, is the mean number
of cells visited by one orbit on a fixed $20\times 20$ family-specific
grid, with the grid ranges given in Appendix~\ref{app:systems}. Its
absolute scale should not be compared between families, whereas
model--truth errors within one family are meaningful. Second, we pool
the section crossings of all 28 orbits into a model cloud $X$ and a
truth cloud $Y$. The symmetric Chamfer
distance \cite{barrow1977chamfer}, the sum of the two directed mean
nearest-neighbor distances between the clouds, is
\begin{equation}
d_{\mathrm{Ch}}(X, Y)
= \frac{1}{|X|}\sum_{u \in X}\min_{v \in Y}\lVert u - v\rVert_2 + \frac{1}{|Y|}\sum_{v \in Y}\min_{u \in X}\lVert v - u\rVert_2.
\label{eq:chamfer}
\end{equation}
Occupancy measures the two-dimensional area covered on the section
and is bounded above by the retained crossing count per orbit, while $d_{\mathrm{Ch}}$ detects displacement or deformation of the section geometry.

\subsection{Finite-time Lyapunov exponents}
\label{sec:lyapunov}

Orbit-wise finite-time largest Lyapunov exponents are computed with the
standard two-trajectory Benettin procedure \cite{benettin1980}, in which
a companion orbit, displaced from the reference orbit by
$d_0 = 10^{-5}$, is advanced alongside it and the separation
$\delta z_n$ is renormalized to $d_0$ after every integration step of size $\Delta t$ (Table~\ref{tab:systems}). The exponent is
the time-averaged logarithmic growth rate of this separation,
\begin{equation}
\lambda_T = \frac{1}{T}\sum_{n=0}^{T/\Delta t - 1}
\log\!\left(\frac{\lVert \delta z_{n+1}\rVert_2}{d_0}\right),
\label{eq:ftle}
\end{equation}
with $T = 240$ unless noted. Truth and every learned field use the same
initial conditions and perturbation directions. We retain $\lambda_T$ as
a continuous finite-time diagnostic and do not turn it into a
fixed-threshold chaos label. Chaotic motion is identified by elevated
$\lambda_T$ together with the section geometry.

\begin{figure}[t]
\centering
\includegraphics[width=\linewidth]{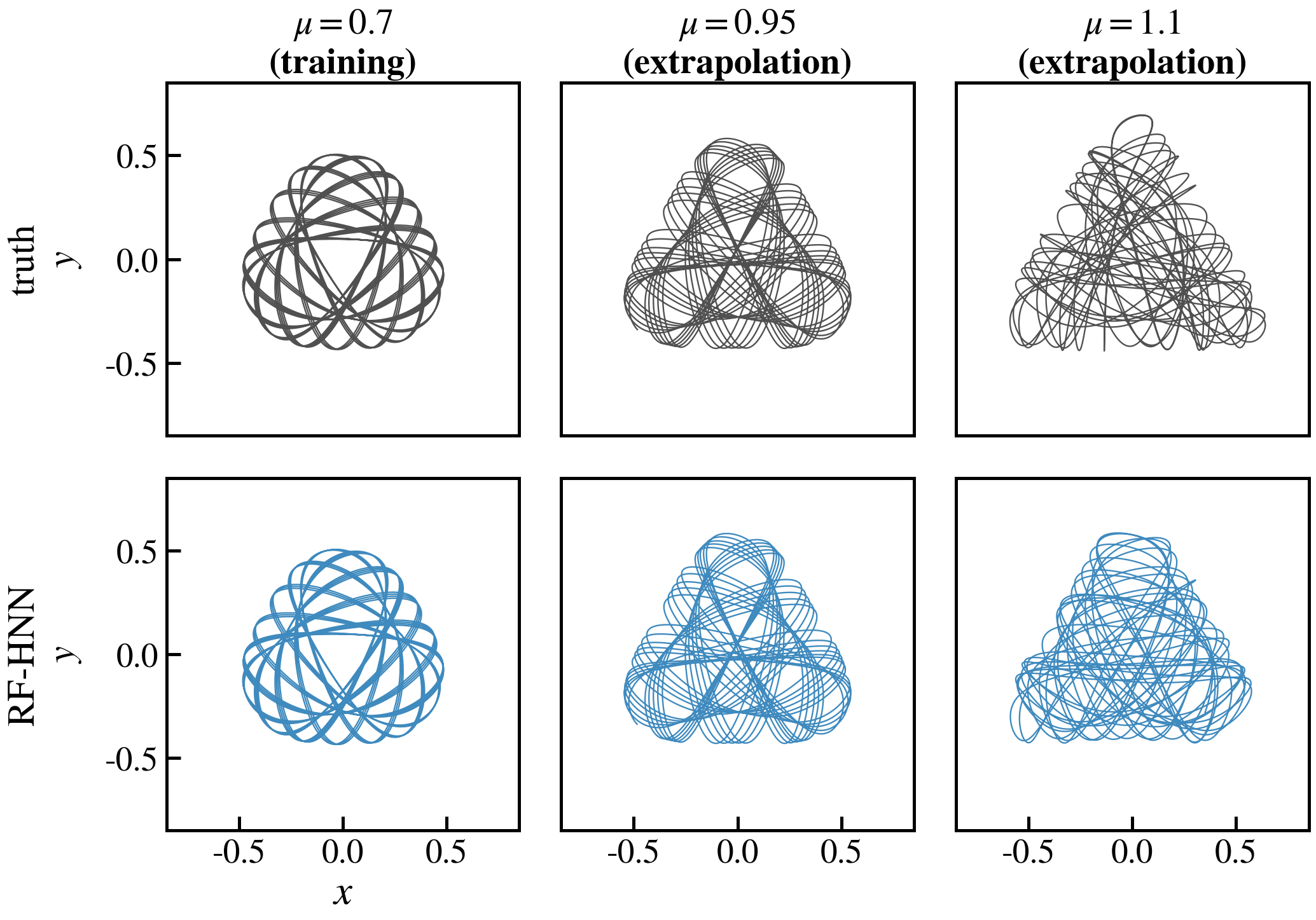}
\caption{Canonical linear-control H\'enon--Heiles family, with no confining wall.
Configuration-space trajectories start from the same section initial condition at every
control. The RF-HNN is fitted at regular controls through $\mu=0.7$. The tracked orbit
remains regular through $\mu = 0.95$ and becomes chaotic at $\mu = 1.1$. Chaotic
trajectories decorrelate, so the two rows are compared by their long-time character
rather than point by point. The three-lobed envelope reflects the three-fold symmetry
of the cubic term. Quantitative Poincar\'e comparisons are shown in
Fig.~\ref{fig:canonicalsec} and Appendix~\ref{app:fullcomparison}.}
\label{fig:canonicaltraj}
\end{figure}

\begin{figure*}[t]
\centering
\includegraphics[width=\linewidth]{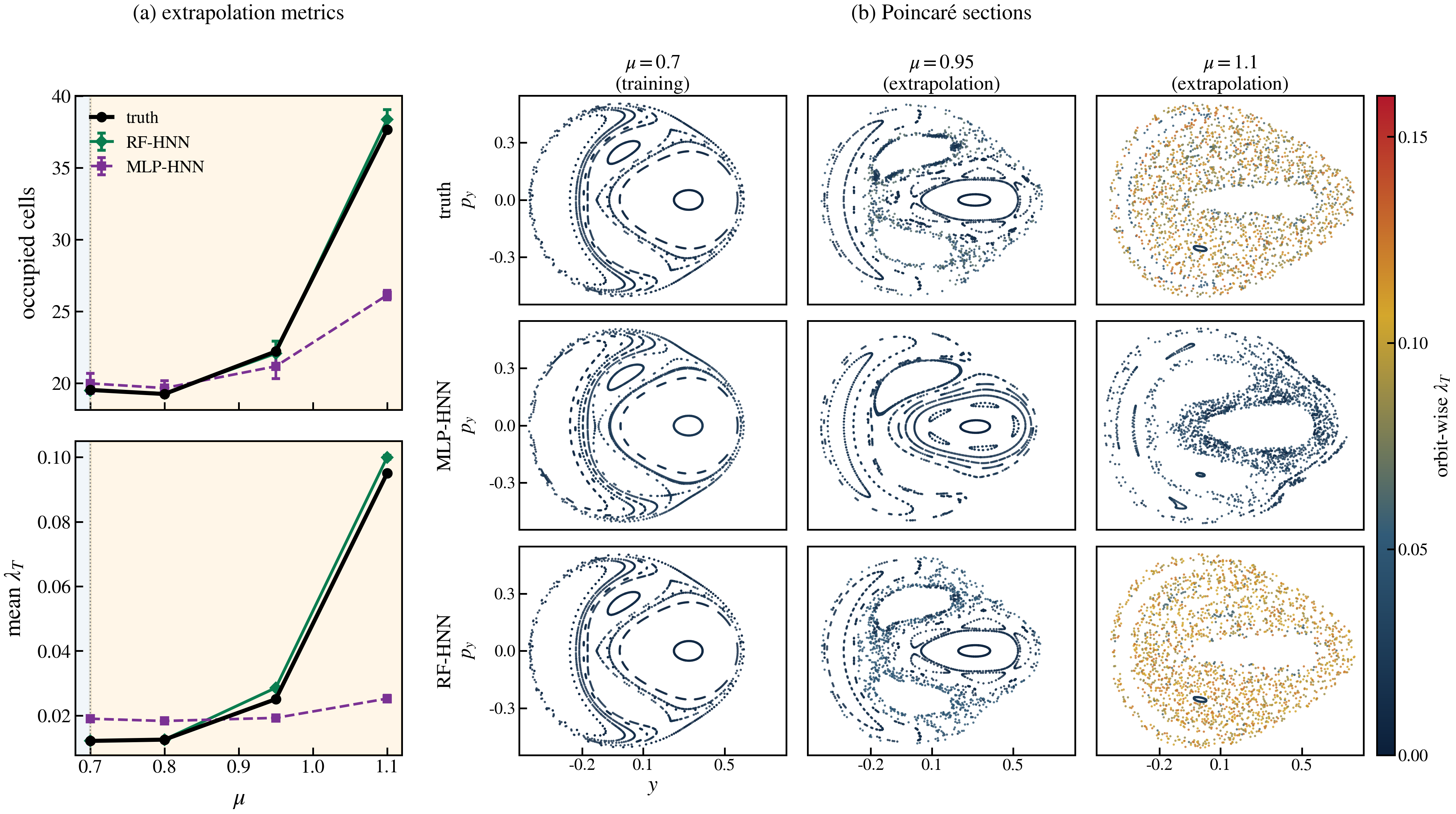}
\caption{Canonical linear-control extrapolation. (a) Occupancy (in cells) and mean
finite-time exponent $\lambda_T$ for truth, the RF-HNN, and the MLP-HNN. The dotted
line marks the last training control, and blue and pale orange backgrounds indicate
the training interval and the extrapolation regime. (b) Poincar\'e sections. Rows show
truth, the MLP-HNN, and the RF-HNN, and columns progress from a training control
($\mu=0.7$) to two extrapolation controls ($\mu=0.95$ and $1.1$). Point color encodes the orbit-wise finite-time exponent
$\lambda_T$ on a common navy-to-red scale. Error bars are one standard deviation over
three independent learned-model realizations. Metric panels use the 28-orbit,
50-crossing sample. Each section panel is a visualization-only cloud of 2400 crossings
from 12 shared initial conditions.}
\label{fig:canonicalsec}
\end{figure*}

\section{Extrapolating the emergence of chaos}
\label{sec:results}

We first test the canonical family, whose potential is unmodified, and
then follow the transition deeper in the bounded family, whose confining
wall permits
extrapolation beyond the escape limit of the canonical potential. Section~\ref{sec:generalization} then extends the comparison to all four
families. All reported occupancy,
Chamfer, and Lyapunov statistics are computed from the 28-orbit sample
of Sec.~\ref{sec:poincare}. Truth and every model are advanced by the
same classical fourth-order Runge--Kutta implementation with the family
time steps $\Delta t$ of Table~\ref{tab:systems}, so that
model-specific time steppers cannot confound the comparison and the
finite-step integration error is common to all flows.

\subsection{The canonical family}
\label{sec:canonicalres}

Figure~\ref{fig:canonicaltraj} tracks a single section initial condition
across the canonical potential of Eq.~\eqref{eq:HH}. At the training
control $\mu = 0.7$ and at $\mu = 0.95$, truth and the RF-HNN trace the same
regular three-lobed geometry. At $\mu = 1.1$ the same orbit turns
chaotic, and the learned flow reproduces this qualitative change,
wandering irregularly over the same three-lobed region as truth, even
though neither extrapolation control was used for training or model
selection. Already at the level of a single orbit, the learned
Hamiltonian therefore continues the family in the control rather than
reproducing a frozen copy of the training flows.

The section statistics support the visual agreement
[Fig.~\ref{fig:canonicalsec}(a); ensemble curves in Fig.~\ref{fig:app-full-comparison}]. At $\mu = 1.1$ the RF-HNN matches the
true occupancy to within about one cell (38.35 versus 37.64), while the
single MLP-HNN and the 20-network ensemble underoccupy the section at
26.13 and 25.00 cells. The finite-time exponents show the same split:
0.100 for the RF-HNN against 0.095 for truth, but 0.025 and 0.024 for
the baselines, values that barely exceed their training-interval level.
The baselines, in other words, prolong the nearly regular dynamics of
the training interval, while the RF-HNN predicts the new instability.
The sections in Fig.~\ref{fig:canonicalsec}(b) show the
same picture. At $\mu = 1.1$, the true section and the RF-HNN fill a
broad chaotic component, while the MLP-HNN leaves substantially more of
the sampled phase space on coherent structures. A broad chaotic
component of this kind appears in no training section, so reproducing it is precisely what the task of Sec.~\ref{sec:task} demands. The canonical test, however, ends while the phase space is
still mixed, with moderate finite-time instability, because the shell
energy $E_0$ reaches the escape threshold near $\mu = 1.13$
(Sec.~\ref{sec:canonicalHH}). The bounded family carries the transition
through to a fully developed chaotic sea
(Secs.~\ref{sec:interp} and~\ref{sec:boundedtrans}).

\begin{figure}[t]
\centering
\includegraphics[width=\linewidth]{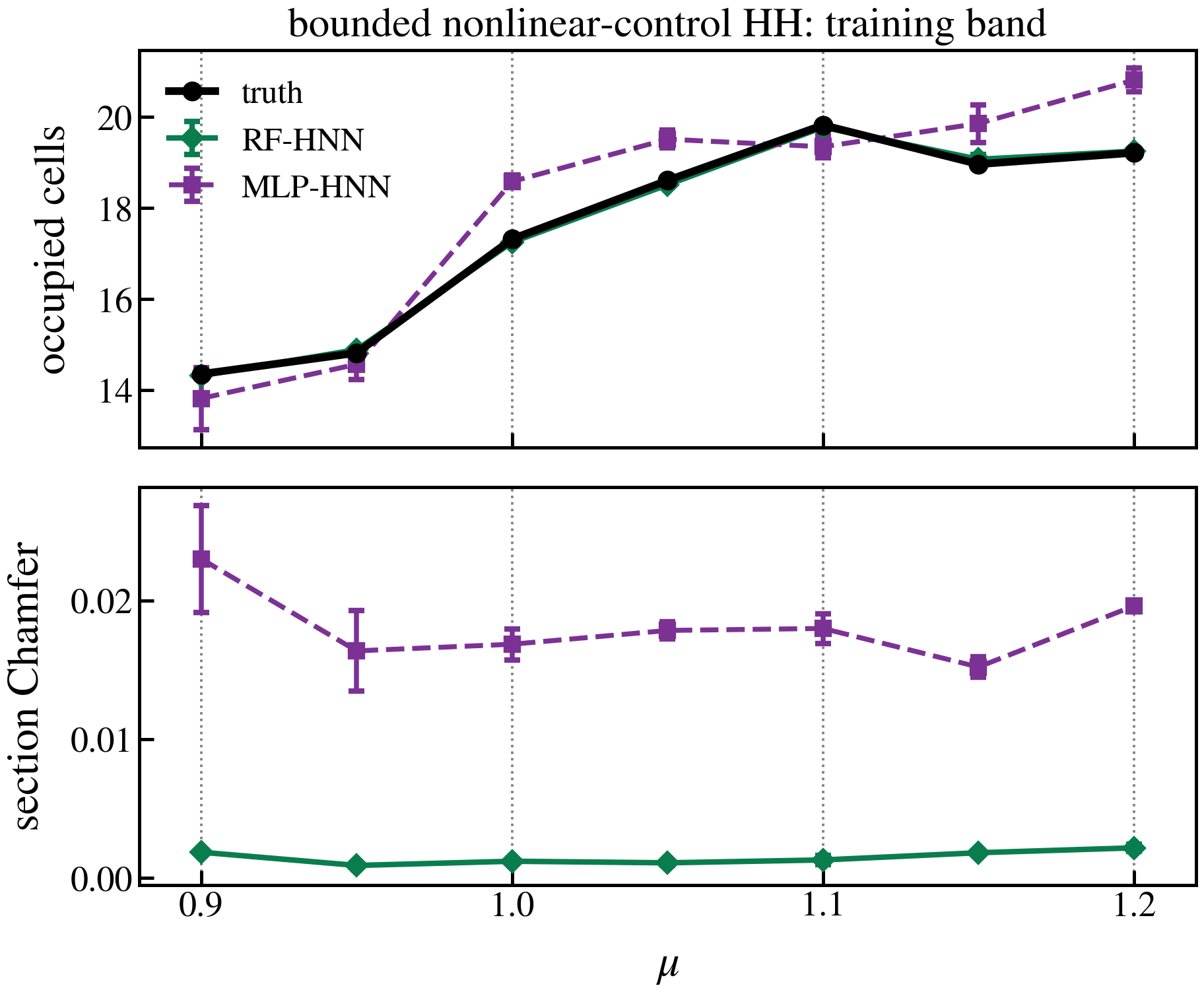}
\caption{Behavior of the bounded nonlinear-control family inside the training
interval: occupancy (top,
in cells) and symmetric section Chamfer distance to truth (bottom), evaluated at the
four training controls and the three interpolation controls. Dotted gray
lines indicate training controls. Curves show truth, the RF-HNN, and the MLP-HNN.
Error bars are one standard deviation over three independent learned-model
realizations.}
\label{fig:interp}
\end{figure}

\begin{figure*}[t]
\centering
\includegraphics[width=\linewidth]{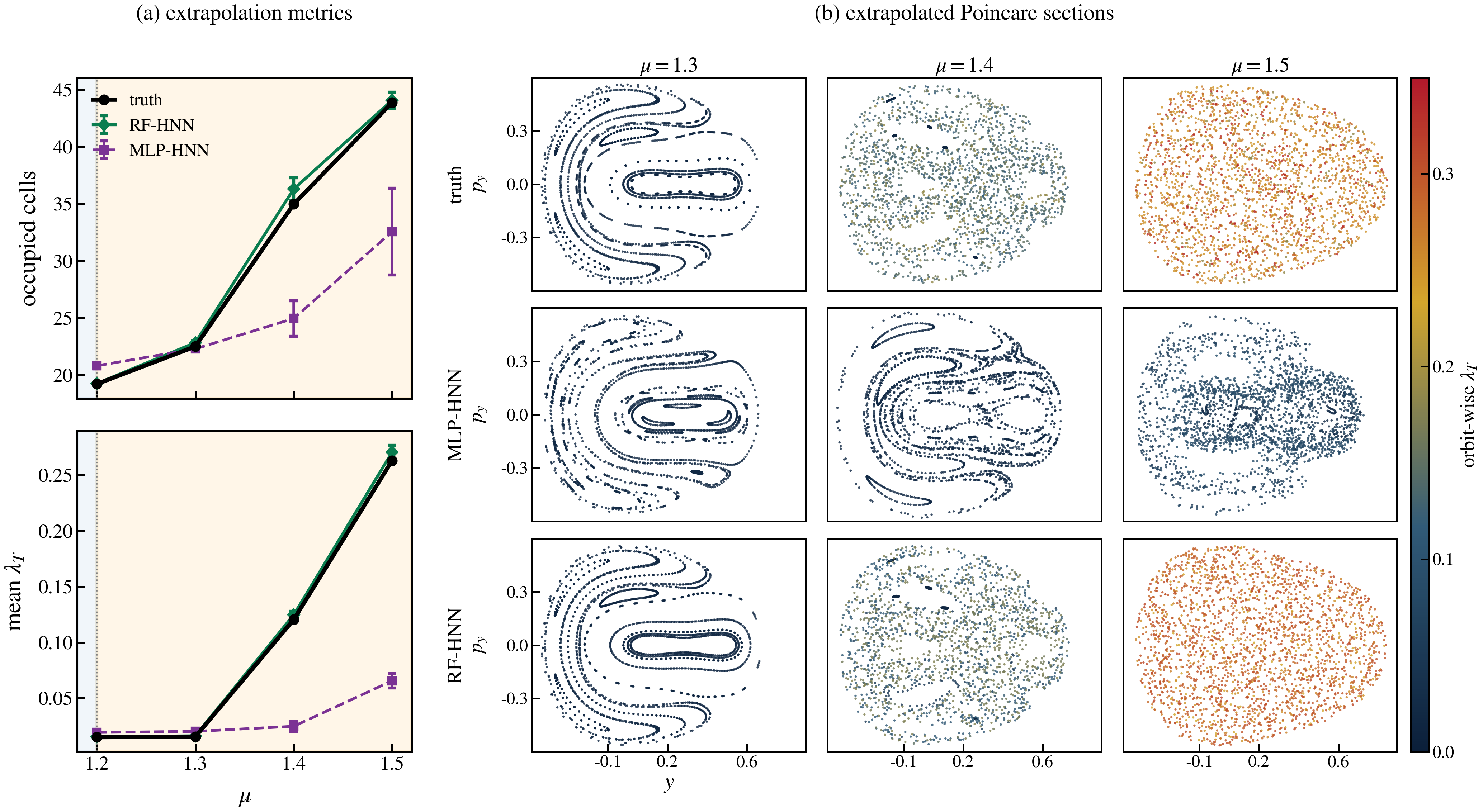}
\caption{Bounded nonlinear-control extrapolation. (a) Occupancy (in cells) and mean
finite-time exponent $\lambda_T$ for truth, the RF-HNN, and the MLP-HNN. The dotted
line marks the last training control, and blue and pale orange backgrounds indicate
the training interval and the extrapolation regime. (b) Rows show truth, the MLP-HNN, and the
RF-HNN. Point color encodes the orbit-wise finite-time exponent $\lambda_T$ on a common
navy-to-red scale. Error bars are one standard deviation over three independent
learned-model realizations. Metric panels use the 28-orbit, 50-crossing sample. Each
section panel is a visualization-only cloud of 2400 crossings from 12 shared initial
conditions.}
\label{fig:boundedext}
\end{figure*}

\subsection{The bounded family within the training interval}
\label{sec:interp}

We now turn to the bounded family of Eq.~\eqref{eq:BHH} and first verify
that a single parameter-aware fit connects the sampled regular shells. Across the
interpolation controls
$\mu = 0.95, 1.05, 1.15$ of the bounded family, the RF-HNN occupancy mean
absolute error is 0.091 cells and the mean section Chamfer distance is
$1.28 \times 10^{-3}$, both at the same small scale as at the training controls
(Fig.~\ref{fig:interp}). This check is not itself evidence of chaos
prediction. It establishes instead that the extrapolation examined next starts from a faithful fit of the regular family, not from a model
that already fails inside the training interval. The corresponding
sections at the interpolation controls are shown in
Appendix~\ref{app:baseline} (Fig.~\ref{fig:interpsec}).

\subsection{The bounded family beyond the training interval}
\label{sec:boundedtrans}

Beyond the last training control $\mu = 1.2$, the occupied section area of the
true family changes little through $\mu = 1.3$ and then grows rapidly
[Fig.~\ref{fig:boundedext}(a)]. Neither the location of this onset nor
the rate of the subsequent growth is supervised, so the flat-then-rising
shape is itself a sharp prediction target. The RF-HNN tracks this
transition quantitatively. At $\mu = 1.4$, truth and the RF-HNN occupy 34.96 and 36.29 cells and yield
$\lambda_T = 0.120$ and 0.125. At $\mu = 1.5$, their occupancies are
43.86 and 44.06, and their exponents are 0.263 and 0.271. The mean RF-HNN
occupancy error over all three extrapolation controls is 0.62 cells.

The sections in Fig.~\ref{fig:boundedext}(b), where each orbit carries
its own finite-time exponent, show that the agreement extends beyond
the summary statistics in~(a). At $\mu = 1.3$, the
learned flow retains the same visible regular structures as truth. At
$\mu = 1.4$, it reproduces the broadened mixed section, and at
$\mu = 1.5$ it fills the same bounded chaotic region as truth. The RF-HNN
thus continues the family of invariant structures beyond the training
interval and reproduces both the onset and the growth of the chaotic
sea, without any data from this regime. The agreement also demonstrates
the nonlinear parameter continuation that the family was designed to
test (Sec.~\ref{sec:boundedHH}), since the network receives $\mu$ while
the underlying Hamiltonian depends on $\mu^2$.

\subsection{Generalization across Hamiltonians}
\label{sec:generalization}

Table~\ref{tab:extrap} extends the test to all four families. The RF-HNN
has the smallest mean occupancy error in every family and the smallest
mean extrapolation Chamfer distance (Appendix~\ref{app:fullcomparison}),
and at the largest extrapolation control $c^*$ its
finite-time exponent is within about 8\% of truth in all four systems. The
full control-dependent curves and error bars are shown in
Appendix~\ref{app:fullcomparison}.

\begin{table*}[t]
\caption{Extrapolation summary. $\mathcal E_{\rm occ}$ is the mean absolute model--truth
occupancy error over all extrapolation controls in Table~\ref{tab:systems}. The four
$\lambda_T^*$ columns give the finite-time exponent ($T=240$) at the largest
extrapolation control $c^*$. Exponents are averaged over 28 orbits and, for RF-HNN and
MLP-HNN statistics, over three independent learned-model realizations. Occupancy is measured on a separate fixed grid for each family.}
\label{tab:extrap}
\centering
\small
\setlength{\tabcolsep}{4.2pt}
\begin{tabular}{@{}lcccccccc@{}}
\toprule
& \multicolumn{3}{c}{$\mathcal E_{\rm occ}$ (cells)}
& $c^*$ & \multicolumn{4}{c}{$\lambda_T^*$} \\
\cmidrule(lr){2-4}\cmidrule(lr){6-9}
Family & RF-HNN & Single & Ensemble & & Truth & RF-HNN & Single & Ensemble \\
\midrule
Linear-$\mu$ H\'enon--Heiles & 0.30 & 4.32 & 4.60 & 1.10 & 0.095 & 0.100 & 0.025 & 0.024 \\
Bounded $\mu^2$ H\'enon--Heiles & 0.62 & 7.18 & 10.02 & 1.50 & 0.263 & 0.271 & 0.066 & 0.057 \\
Confined Barbanis-type & 0.20 & 1.79 & 2.16 & 1.60 & 0.063 & 0.067 & 0.032 & 0.031 \\
Bilinear Morse pair & 0.19 & 1.29 & 1.29 & 1.10 & 0.172 & 0.168 & 0.102 & 0.106 \\
\bottomrule
\end{tabular}
\end{table*}

The same table shows that this extrapolation does not follow from
Hamiltonian structure alone. The MLP-HNN, trained on
the same shells with the same objective,
remains a valid Hamiltonian model and stays regular on the training tori,
yet its extrapolated flows remain too regular once the true chaotic component expands. At the
largest extrapolation control of the bounded family, the single MLP-HNN and the
20-network ensemble occupy 32.55 and 26.86 cells against 43.86 for truth,
with exponents 0.066 and 0.057 against 0.263. The failure is visible in
Fig.~\ref{fig:boundedext}(b). At $\mu = 1.4$ the single network's section
(middle row) retains torus-like invariant curves with uniformly low
exponents where the true section is already a broad mixed sea, and at
$\mu = 1.5$ it spreads over more of the section while its exponents
remain far below truth. The ensemble does not
repair this deficit. Its mean occupancy error matches or exceeds the
single MLP-HNN's in all four families.
Averaging suppresses member-to-member variance, but the extrapolated
Hamiltonians share a bias toward sections that are too regular.

\section{Discussion}\label{sec:discussion}

The main result of this study is that a parameter-aware RF-HNN can extrapolate the emergence of a broad chaotic regime from data dominated by regular motion. No data from the extrapolation regime enter training or model selection, yet the learned Hamiltonian reproduces the breakup of invariant structures and the growth of chaotic components at unseen controls across four Hamiltonian families. Because the flow of the learned scalar Hamiltonian is integrated autonomously, the prediction is not a direct classification of chaos or a short-time extension of observed trajectories. The out-of-range continuation of the learned Hamiltonian generates the long-time phase-space organization and instability of the new regime.

Evaluating this prediction requires measures that survive the rapid decorrelation of chaotic trajectories. We therefore use Poincar\'e-section occupancy, section geometry, and finite-time Lyapunov exponents, rather than long-horizon pointwise error, to probe the extent, location, and instability of the learned flow. Their combined agreement is important because the transition appears differently across the systems, with some showing a strong change in occupied area and others a clearer increase in instability. The consistent RF-HNN behavior across different potentials, symmetries, and couplings indicates that the result is not tied to one H\'enon--Heiles parameterization.

The contrast with the MLP-HNN baselines exposes an ambiguity specific to parameter extrapolation. Matching the field $J\nabla_z H$ on energy shells at a finite set of controls constrains the vector field where data are available, but it does not uniquely determine its continuation in $c$. Distinct Hamiltonian models can therefore reproduce the observed regular regime while generating different phase-space organizations outside it. This is what the comparison shows. The MLP-HNNs reproduce the regular in-band behavior (inside the training interval), while their continuations delay or suppress the breakup of invariant structures and the associated growth of instability. The ensemble does not repair this discrepancy because averaging reduces member-to-member variance, not a bias shared by the learned continuations. The relevant distinction is therefore not only between Hamiltonian and non-Hamiltonian learning, but among Hamiltonian continuations that are nearly indistinguishable on the observed interval.

The mechanism by which the RF-HNN selects the more faithful continuation is not yet isolated, but the results suggest a specific role for its learning procedure. Once the random features are fixed, fitting is a regularized linear inverse problem, and the feature scales set how rapidly the representation can vary in state and control. In-band validation over these scales may therefore favor a lower-complexity continuation in $c$ without any access to the target regime, and the convex readout makes this selection computationally practical. The exploratory tests in Appendix~\ref{app:mlpsweep} disfavor the simplest alternatives: increasing the MLP-HNN width, changing its activation, or refitting its final layer by ridge regression does not systematically recover the transition. These tests do not identify a unique mechanism, however. Matched ablations that equalize in-band approximation error and effective capacity will be needed to separate the roles of the fixed representation, the control-channel scaling, the regularized readout, and the hyperparameter selection.

A second implication is that the predictability of a dynamical regime can survive the unpredictability of its individual trajectories. Beyond the transition, pointwise rollouts decorrelate rapidly, yet the RF-HNN reproduces the occupied region of the section, the surviving regular structures, and the finite-time instability of the extrapolated flow. Vector-field data from a regular-dominated regime can thus contain enough information to extrapolate the parameter dependence that governs later resonance overlap, provided that the learning procedure selects an appropriate continuation. The claim is consequently about long-time phase-space organization, not about pointwise forecasting or global recovery of the Hamiltonian.

The present study establishes these results in a controlled setting:
synthetic two-degree-of-freedom families, one smoothly varying control,
clean vector-field data on a single energy shell, and moderate
extrapolation distances. Extending the analysis to noisy and sparsely
sampled trajectories, multiple energies or controls, and
higher-dimensional Hamiltonian systems will determine how broadly the
result persists. A further challenge is to decide, from in-band
information alone, when an extrapolated phase portrait should be
trusted, because a small in-band field error does not by itself certify
the extrapolated dynamics.

The central conclusion is that matching the Hamiltonian vector field at
the observed controls and reproducing the phase-space transition of the
underlying family beyond them are distinct requirements. The former can
be optimized and validated inside the observed interval, whereas the
latter depends on how the fitted Hamiltonian continues in the control
beyond it. In the four families studied here, the RF-HNN meets both
requirements, generating the breakup of invariant structures and the
growth of a chaotic sea absent from training and model selection.

\begin{acknowledgments}
J. Choi was supported by a KIAS Individual Grant (No. AP092902) via the Center for
Artificial Intelligence and Natural Sciences at the Korea Institute for Advanced Study
(KIAS). This work was supported by the Center for Advanced Computation at KIAS.
\end{acknowledgments}

\section*{Data availability}
The code and data that support the findings of this study are
available from the corresponding author upon reasonable request.

\appendix

\section{Additional systems and implementation details}\label{app:systems}

\begin{table*}[!t]
\caption{Selected depth-2 RF-HNN hyperparameters. $s_z$ and $s_c$ scale the state
weights $W_z$ and the control weights $w_c$, and $g_2/\sqrt{N}$ scales the
second-layer matrix $W$.}
\label{tab:hyperparameters}
\centering
\small
\setlength{\tabcolsep}{5pt}
\begin{tabular}{@{}lccccccc@{}}
\toprule
Family & Activation & $N$ & $s_z$ & $s_c$ & $g_2$ & $\log_{10}\alpha$ & Val. control \\
\midrule
Linear HH & GELU & 2000 & 0.416 & 0.0266 & 2.415 & $-10.363$ & 0.6 \\
Bounded HH & GELU & 1800 & 0.378 & 0.0421 & 1.043 & $-10.086$ & 1.1 \\
Conf. Barbanis & softplus & 1800 & 0.373 & 0.0201 & 0.0340 & $-12.013$ & 1.0 \\
Morse pair & softplus & 2000 & 0.865 & 0.0203 & 0.888 & $-8.946$ & 0.4 \\
\bottomrule
\end{tabular}
\end{table*}

The original Barbanis resonance model combines harmonic terms with a cubic $xy^2$ coupling
and contains no quartic wall \cite{barbanis1966}. Our auxiliary family is instead the
quartically confined Barbanis-type Hamiltonian
\begin{equation}
\begin{aligned}
H^{\mathrm{B}}_{c}(z)
&= \tfrac{1}{2}\bigl(p_x^2+p_y^2\bigr)
 + \tfrac{1}{2}\bigl(x^2+y^2\bigr) \\
&\quad + c\,x y^2 + \kappa_{\mathrm{B}}\bigl(x^2+y^2\bigr)^2,
\end{aligned}
\label{eq:barbanis}
\end{equation}
with $\kappa_{\mathrm{B}} = 0.3$ and $E_0=0.18$. The added quartic term makes this a globally bounded variant rather than
the standard Barbanis system. Relative to the H\'enon--Heiles families, the $xy^2$ coupling changes
both the symmetry and the resonant structure, while the quartic
confinement is retained.

The onsite Morse potential is canonical \cite{morse1929}, but there is no unique standard
coupling of two Morse oscillators. Molecular coupled-Morse models may use, for example, a
momentum coupling $G_{12}p_1p_2$ \cite{kryvohuz2010}. Here we use the deliberately simple
bilinearly coupled Morse pair
\begin{equation}
\begin{aligned}
H^{\mathrm{M}}_{c}(z)
&= \tfrac{1}{2}\bigl(p_x^2+p_y^2\bigr)
 + \bigl(1-e^{-x}\bigr)^2 \\
&\quad + \bigl(1-e^{-y}\bigr)^2 + c\,x y,
\end{aligned}
\label{eq:morse}
\end{equation}
also at $E_0=0.18$. Each uncoupled Morse oscillator is integrable, so the mixed dynamics
arises entirely from the bilinear coupling, which breaks separability. Because the coupling does not confine
the potential globally, escape is possible in principle, and the family therefore serves
only as a supplementary test at the fixed shell energy and finite integration times used
here. In practice, every orbit evaluated in the main comparison remains in the potential well from which the data are sampled and completes its 50 section crossings without escaping.

The configuration-space proposal boxes used to sample training shells are
$[-0.7,0.7]\times[-0.7,0.9]$ for linear H\'enon--Heiles, $[-0.9,0.9]^2$ for bounded
H\'enon--Heiles, $[-1.1,1.1]^2$ for the confined Barbanis-type model, and
$[-0.8,1.5]^2$ for the Morse pair. The affine
control normalizations supplied to the MLP-HNN, which map each family's
control values to a common order-one range, are $(\mu-0.5)/0.6$,
$(\mu-0.8)/0.5$, $(c-1.0)/0.6$, and $(c-0.6)/0.4$, respectively.

For occupancy, the fixed $(y,p_y)$ ranges are $[-0.7,0.9]\times[-0.6,0.6]$ for linear
H\'enon--Heiles, $[-0.8,0.8]\times[-0.7,0.7]$ for bounded H\'enon--Heiles and the Morse pair,
and $[-1.1,1.1]^2$ for the confined Barbanis-type model. These ranges and the
$20\times20$ resolution are held fixed
across truth and all models within a family.

The MLP-HNN is trained with Adam (initial learning rate
$2\times10^{-3}$, no weight decay, cosine schedule) on the same field-matching objective
and control set as the RF-HNN, using 12000 shell states per control, minibatches of 2048,
and 12000--14000 updates depending on the family. States are standardized by the
training-set mean and standard deviation before entering the network. Twenty independent
networks are trained per family. The single MLP-HNN statistics average the first three members (training seeds 100--102), fixed by the protocol before any evaluation, and the
ensemble prediction is the gradient of the mean of all 20 learned
Hamiltonians.

Halving the time step, varying the perturbation size over
$d_0 \in \{10^{-4}, 10^{-5}, 10^{-6}\}$, and computing the exponents by
direct integration of the tangent equation all lead to the same
conclusions. Integrations of
truth extended to $T = 3840$ confirm decaying finite-time instability
at every training control.

\section{Hyperparameter optimization}\label{app:hpo}

The RF-HNN uses two hidden layers of common width $N$. We run a
separate 100-trial Bayesian-optimization study with
Optuna \cite{optuna2019} for each activation in
$\{\tanh,\mathrm{GELU},\mathrm{softplus}\}$ and each Hamiltonian family. The width candidates are $N=200,400,\ldots,2000$. The log-uniform
ranges are $s_z\in[0.05,4]$, $s_c\in[0.02,6]$, and $g_2\in[0.03,4]$, while
$\log_{10}\alpha\in[-16,-2]$. Bias scales are fixed at one, with every component sampled
independently from $[-1,1]$, and the frozen matrices are dense. GELU denotes the exact
$x\Phi(x)$ activation, where $\Phi$ is the standard normal cumulative distribution function. Each trial fits 8000 shell states per fitted control and is validated
on 3000 independently sampled states.

The trial fits use the training controls $\{0.4,0.5,0.7\}$,
$\{0.9,1.0,1.2\}$, $\{0.8,1.2\}$, and $\{0.3,0.5\}$ for linear HH,
bounded HH, Barbanis, and Morse, with the remaining training control
of each family ($0.6$, $1.1$, $1.0$, and $0.4$, respectively) used for
validation. Thus every trial and the activation choice use only the designated
training interval. All search trials share one fixed random-feature
seed. The selected configuration is then refitted with three fresh
feature seeds that were never used during the search, and reported
results average these three models. Table~\ref{tab:hyperparameters}
gives the selected configurations.

\section{Architecture sweep of the MLP-HNN baseline}\label{app:mlpsweep}

Table~\ref{tab:mlpsweep} reports the exploratory architecture sweep
referenced in Sec.~\ref{sec:discussion}: widths 200, 1000, and 2000
combined with tanh, GELU, and softplus activations, trained with the
otherwise unchanged MLP-HNN protocol. No configuration reaches the truth-level exponents at the largest extrapolation control, the width-2000 tanh and GELU variants destabilize the Morse family, and the width-1000 tanh and GELU variants show partial escape there. In an additional test
not included in Table~\ref{tab:mlpsweep}, we refitted the final layer of
each trained network by the same ridge regression used for the RF-HNN
readout. This refit recovered the transition only in a minority of
configurations, and no in-band criterion identified those configurations
in advance. These results indicate that the gap reflects the full
learning procedure rather than the readout alone, though we do not claim
that conventionally trained HNNs are intrinsically unable to extrapolate.

\begin{table*}[t]
\caption{Architecture sweep of the back-propagated MLP-HNN baseline. Every model keeps
the unchanged training protocol (12000 states per training control, Adam at
$2\times10^{-3}$, cosine schedule, family step counts, training seeds 100--102) and
varies only the hidden width and activation. GELU and softplus give the MLP-HNN the
activation family of the selected RF-HNN trunks, and widths 1800--2000 are the selected
trunk widths. $\mathcal E_{\rm occ}$ and $\lambda_T^*$ are defined as in
Table~\ref{tab:extrap}. Entries are means over the three members. ``esc.''\ marks
configurations for which escape of extrapolated orbits from the confining region prevents
the statistics in some or all members. The dagger marks entries that include members with
partial escape or incomplete sections at the largest extrapolation control. RF-HNN and
truth rows are repeated for reference.}
\label{tab:mlpsweep}
\centering
\small
\setlength{\tabcolsep}{3.6pt}
\begin{tabular}{@{}llcccccccc@{}}
\toprule
& & \multicolumn{2}{c}{Linear-$\mu$ HH} & \multicolumn{2}{c}{Bounded $\mu^2$ HH}
& \multicolumn{2}{c}{Conf.\ Barbanis} & \multicolumn{2}{c}{Morse pair} \\
\cmidrule(lr){3-4}\cmidrule(lr){5-6}\cmidrule(lr){7-8}\cmidrule(lr){9-10}
Width & Act. & $\mathcal E_{\rm occ}$ & $\lambda_T^*$ & $\mathcal E_{\rm occ}$ & $\lambda_T^*$
& $\mathcal E_{\rm occ}$ & $\lambda_T^*$ & $\mathcal E_{\rm occ}$ & $\lambda_T^*$ \\
\midrule
200  & tanh     & 4.32 & 0.025 & 7.18 & 0.066 & 1.79 & 0.032 & 1.29 & 0.102 \\
200  & GELU     & 2.68 & 0.039 & 4.21 & 0.135 & 0.90 & 0.042 & 5.22 & 0.054 \\
200  & softplus & 2.44 & 0.050 & 3.23 & 0.126 & 0.40 & 0.052 & 1.07 & 0.103 \\
1000 & tanh     & 3.90 & 0.027 & 5.72 & 0.063 & 1.18 & 0.031 & 6.11$^\dagger$ & 0.047 \\
1000 & GELU     & 2.25 & 0.047 & 2.73 & 0.159 & 1.12 & 0.039 & 3.82$^\dagger$ & 0.084 \\
1000 & softplus & 2.32 & 0.053 & 2.18 & 0.101 & 0.86 & 0.030 & 4.70 & 0.049 \\
2000 & tanh     & 3.91 & 0.030 & 5.19 & 0.061 & 1.21 & 0.039 & \multicolumn{2}{c}{esc.} \\
2000 & GELU     & 3.69 & 0.028 & 2.28 & 0.125 & 1.26 & 0.028 & \multicolumn{2}{c}{esc.} \\
2000 & softplus & 2.93 & 0.033 & 4.10 & 0.095 & 0.99 & 0.026 & 4.98 & 0.041 \\
\midrule
\multicolumn{2}{@{}l}{RF-HNN}      & 0.30 & 0.100 & 0.62 & 0.271 & 0.20 & 0.067 & 0.19 & 0.168 \\
\multicolumn{2}{@{}l}{Truth ($\lambda_T^*$)} & --- & 0.095 & --- & 0.263 & --- & 0.063 & --- & 0.172 \\
\bottomrule
\end{tabular}
\end{table*}

\section{Poincar\'e-section geometry at the interpolation controls}\label{app:baseline}

\begin{figure}[!htb]
\centering
\includegraphics[width=\linewidth]{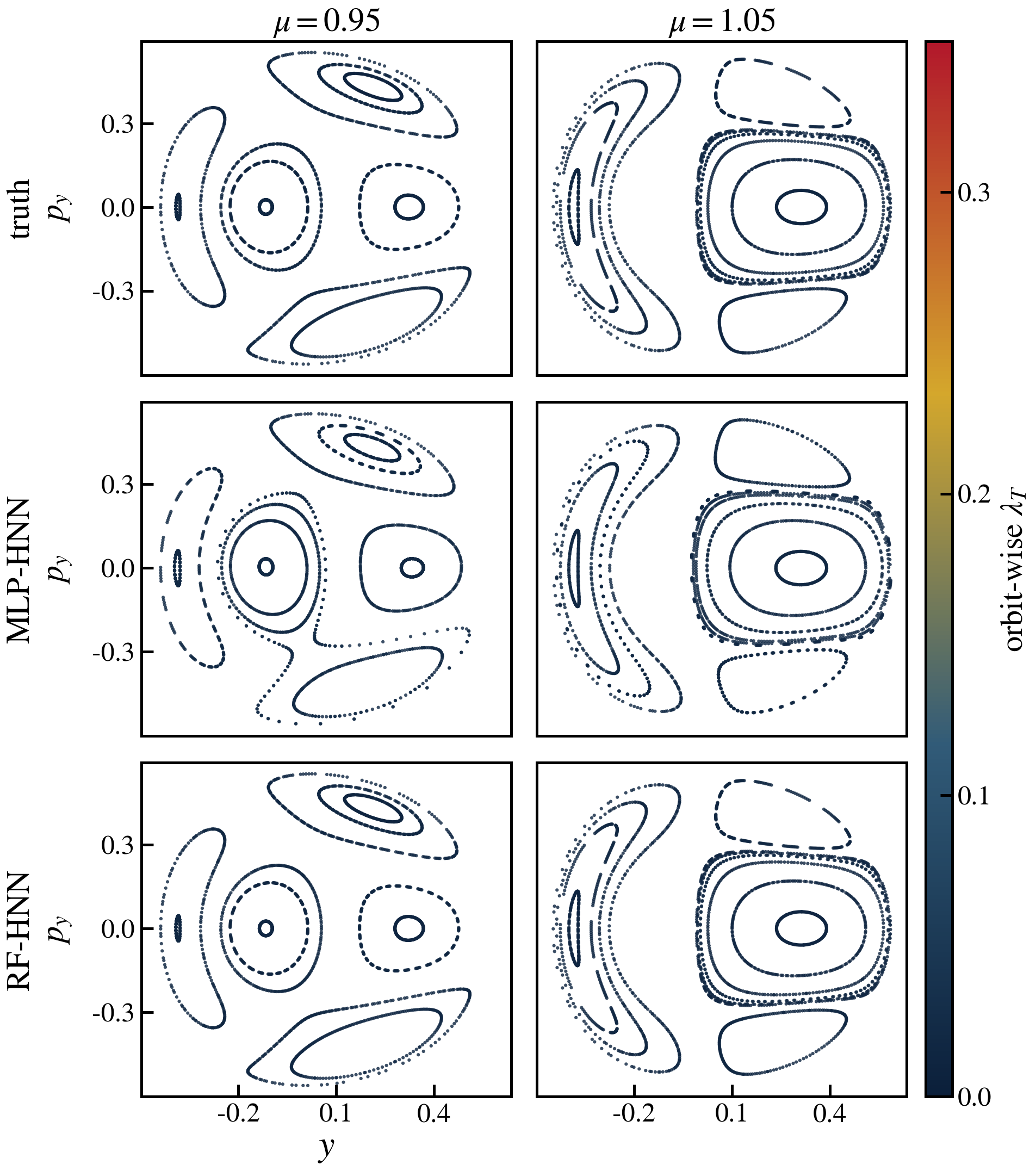}
\caption{Interpolation counterpart of Fig.~\ref{fig:boundedext} for the bounded
nonlinear-control family. Rows show truth, the MLP-HNN, and the RF-HNN
at the interpolation controls $\mu = 0.95$ and $1.05$. Point color encodes the orbit-wise
finite-time exponent $\lambda_T$ on the same navy-to-red scale as
Fig.~\ref{fig:boundedext}. The uniformly dark panels reflect that every flow remains regular
here. Each visualization-only panel contains 2400
crossings from 12 shared initial conditions. Reported statistics use the separate 28-orbit
sample.}
\label{fig:interpsec}
\end{figure}

Figure~\ref{fig:interpsec} shows the sections behind the
interpolation check of Sec.~\ref{sec:interp}, at the interpolation controls
$\mu = 0.95$ and $1.05$.

 Inside the training interval both learned
models stay close to truth. They match the true occupancy to within one
cell, every section remains uniformly low-$\lambda_T$ on the color
scale shared with Fig.~\ref{fig:boundedext}, and at $\mu = 1.05$ the
invariant curves of both models are nearly indistinguishable from the
true ones. At $\mu = 0.95$ the MLP-HNN curves show a slight
displacement from the true curves. The Chamfer distance quantifies the
residual difference. At the two displayed controls, the RF-HNN Chamfer distance is about seventeen times smaller ($0.0009$--$0.0011$ against $0.016$--$0.018$ for the MLP-HNN). The much
larger separation in occupancy appears only beyond the last training
control [Fig.~\ref{fig:boundedext}].

\section{Full four-family comparison}\label{app:fullcomparison}

\begin{figure*}[!t]
\centering
\includegraphics[width=\linewidth]{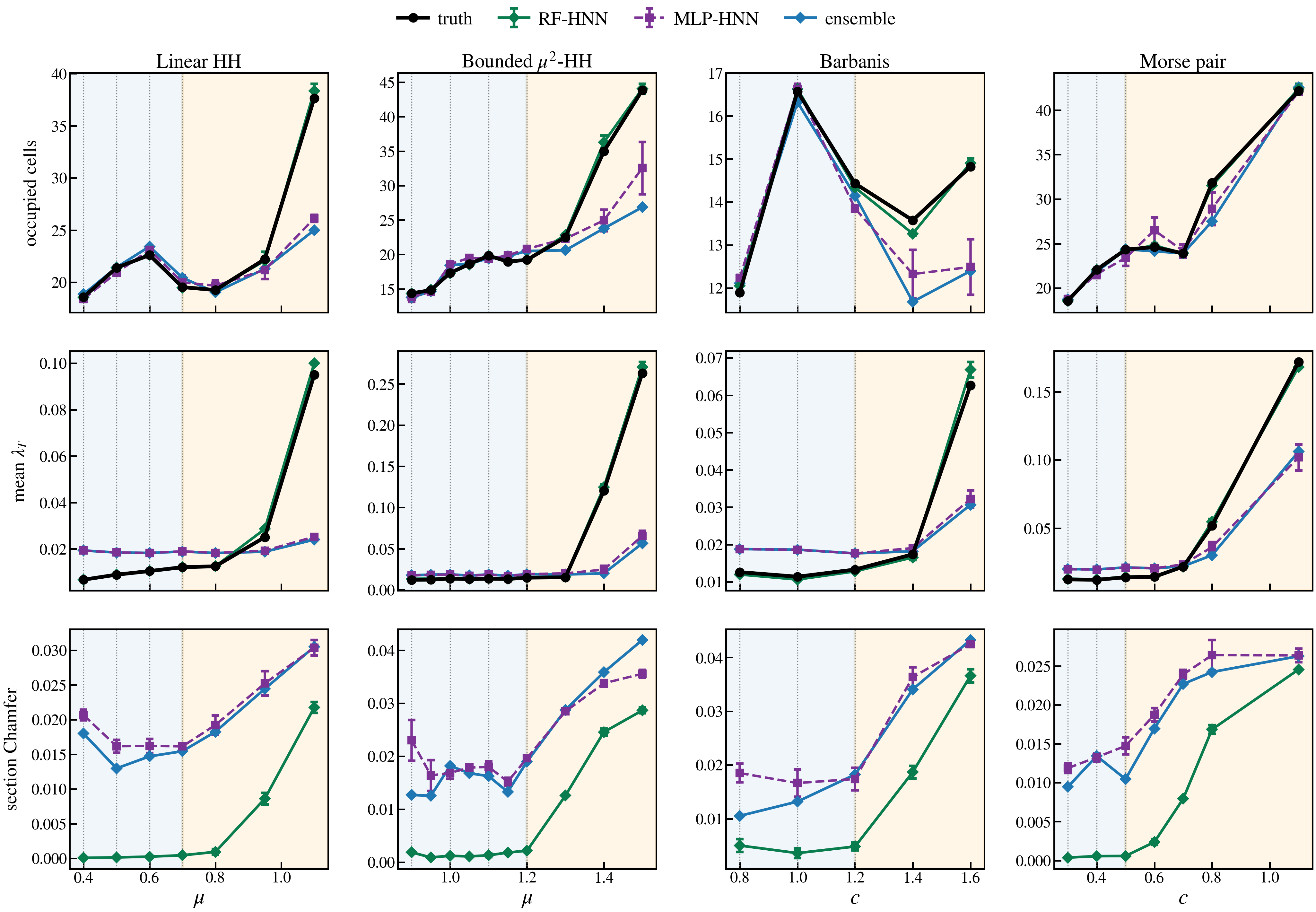}
\caption{Full comparison for linear H\'enon--Heiles, bounded nonlinear-control
H\'enon--Heiles, the confined Barbanis-type model, and the bilinear Morse pair. Rows show
mean occupied cells per orbit, mean finite-time exponent $\lambda_T$, and symmetric
section Chamfer distance to truth. Dotted vertical lines mark the training controls. Blue and
pale orange backgrounds indicate the training interval and the extrapolation regime. Error bars
are one standard deviation over three independent RF-HNN or MLP-HNN realizations.}
\label{fig:app-full-comparison}
\end{figure*}

Figure~\ref{fig:app-full-comparison} gives the control-dependent occupancy, finite-time-exponent,
and Chamfer curves behind Table~\ref{tab:extrap}. Inside the training interval the occupancy curves
remain close to truth, while the Chamfer curves already separate the
models, consistent with Fig.~\ref{fig:interpsec}. The curves diverge beyond the last
training control, and the manner of the divergence differs by family. In the canonical family the
baselines stay close through $\mu=0.95$ but fall sharply behind by $\mu=1.1$, the ensemble
more strongly than the single MLP-HNN. In the bounded family the occupancy
changes little through $\mu=1.3$ and then rises steeply. Both baselines
follow the initial segment but not the rise, so their error appears exactly where the
qualitatively new growth appears.

The two auxiliary families refine rather than change this picture. In the confined
Barbanis-type model the quartic wall keeps the occupied area within a narrow range, so the
transition expresses itself less in occupancy than in the exponents. Both
baselines underoccupy the section at the largest extrapolation control, while their exponents fall to about
half the true value. In the Morse pair the occupancy of both baselines lies close to truth
at the largest extrapolation control, while their exponents remain well
below truth, so occupancy alone would understate the remaining
difference there. Across all four families
the RF-HNN Chamfer distance stays at or below that of both baselines at every extrapolation control, and its occupancy tracks the true curves throughout, including the non-monotonic segment of the Barbanis-type
family.

\clearpage
\bibliography{ms}

\end{document}